\documentclass[prl,aps,showpacs,twocolumn,superscriptaddress,reprint]{revtex4-1}

\usepackage{rotating}	
\usepackage{amsmath}	
\usepackage{amssymb}	
\usepackage{mathtools}
\usepackage{color}
\usepackage[hidelinks]{hyperref}
\hypersetup{  
colorlinks=true,
citecolor=blue} 
\usepackage{bm}
\usepackage{textcomp}
\usepackage{lmodern}

\newcommand{\f}[2]{
		\mathchoice%
			{\dfrac{#1}{#2}}
	    	{\dfrac{#1}{#2}}
			{\frac{#1}{#2}}
			{\frac{#1}{#2}}
}

\newcommand{\dd}{\mathrm{d}}

\newcommand{\ddf}[3][]{%
        \ifthenelse{\equal{#1}{}}{%
                \ensuremath{\f{\dd#2}{\dd#3}}%
        }{%
                \ensuremath{\f{\dd^{#1}#2}{\dd{#3}^{#1}}}%
        }%
}
\newcommand{\Dp}[3][]{
        \ifthenelse{\equal{#1}{}}{%
                \ensuremath{\f{\partial#2}{\partial#3}}%
        }{%
                \ensuremath{\f{\partial^{#1}#2}{\partial{#3}^{#1}}}%
        }%
}

\newcommand{\MF}{\textsc{mf}}

\newcommand{\beq}{ \begin{equation} }
\newcommand{\eeq}{ \end{equation} } 
\newcommand{\bea}{ \begin{eqnarray} }
\newcommand{\eea}{ \end{eqnarray} }

\begin{document}

\title{Collective effects enhancing power and efficiency}

\author{Hadrien Vroylandt}
\affiliation{Laboratoire de Physique Th\'eorique (UMR8627), CNRS, Univ. Paris-Sud, Universit\'e Paris-Saclay, 91405 Orsay, France}
\author{Massimiliano Esposito}
\affiliation{Complex Systems and Statistical Mechanics, Physics and Material Science Research Unit, University of Luxembourg, L-1511 Luxembourg, G.D. Luxembourg}
\author{Gatien Verley}
\affiliation{Laboratoire de Physique Th\'eorique (UMR8627), CNRS, Univ. Paris-Sud, Universit\'e Paris-Saclay, 91405 Orsay, France}

\date{\today}

\begin{abstract}
Energy conversion is most efficient for micro or nano machines with tight coupling between input and output power. To reach meaningful amounts of power, ensembles of $N$ such machines must be considered. We use a model system to demonstrate that interactions between $N$ tightly coupled nanomachines can enhance the power output per machine. Furthermore, while interactions break tight coupling and thus lower efficiency in finite ensembles, the macroscopic limit ($N \rightarrow \infty$) restores it and enhances both the efficiency and the output power per nanomachine.
\end{abstract}

\maketitle


\emph{Introduction---} Improving the performances of machines at the macroscopic scale has always been a central objective of thermodynamics \cite{Callen1985_vol, Book_Bejan2006}. 
Recent investigations have shown that by operating at small-scales, high efficiencies can be reached e.g. for thermoelectric devices \cite{Humphrey2002_vol89, Humphrey2005_vol94, Dresselhaus2007_vol19, Whitney2014_vol112, Benenti2017_vol694}, photoelectric cells \cite{Rutten2009_vol80}, or molecular motors \cite{Julicher1997_vol69, Gaspard2007_vol247, Lau2007_vol99, Parmeggiani1999_vol60, Lipowsky2010_vol42, Altaner2015_vol92}. 
An important ingredient in this regard is the property of tight coupling. Close to equilibrium, this property implies that the Onsager matrix which characterizes how the input and output dissipative flows are couped to each others becomes singular. Away from equilibrium, it implies that every cyclic processes performed by the machine carries the input flow as well as the output flow in the same proportion. In other words, the input and output flows are completely correlated and their ratio does not fluctuate \cite{Polettini2015_vol114}.
Tight coupling is most naturally fulfilled in very small devices described by stochastic networks containing a single cycle \cite{Seifert2012_vol75}. It is known to lead to higher efficiencies both close \cite{Julicher1997_vol69, Parmeggiani1999_vol60, Entin-Wohlman2014_vol89} and far form equilibrium such as at maximum power \cite{Seifert2012_vol75, VandenBroeck2005_vol95, Esposito2009_vol102, Park2016_vol94}. 

Despite extensive studies on the power-efficiency trade-off \cite{Shiraishi2016_vol117, Proesmans2016_vol116, Pietzonka2017_vol} and growing evidence that reversible efficiencies may be approached away from equilibrium \cite{Esposito2009_vol85, Seifert2011_vol106, Lee2016_vol, Polettini2015_vol114, Polettini2016_vol,Campisi2016_vol7, Koning2016_vol89}, the drawback of nano-machines remains the low power they deliver. A natural way to overcome this limitation is to assemble large numbers of nano-machines \cite{Juelicher1995_vol75}. This immediately raises the question whether interactions amongst those machines may be used to improve the performance per machine. This is a-priori not obvious because interactions are expected to decorrelate the input and output flows and to thus brake the tight coupling property. While mean field treatments in the context of molecular motors and coupled oscillators have demonstrated the existence of such cooperative effects \cite{Golubeva2013_vol88, Golubeva2012_vol109, Imparato2015_vol17}, little is known on their dependence in the number of machines. 

Our aim in this letter is to study the efficiency and output power of a collective machine made of $N$ interacting unicyclic nanomachines, focusing on the role of the interaction strength and of $N$. The machines are two level systems which repel each other when in different states and which are subjected to a nonconservative force $F$ and in contact with two thermal reservoirs at inverse temperatures $\beta_\nu = 1/(k_B T_\nu)$, with $\nu=1,2$, $k_B =1$ and $\beta_1 > \beta_2$. A variant of this machine was introduced in Ref.~\cite{Cleuren2001_vol54} to study negative mobility. It is simple enough to solve the mean field theory exactly which reveals a pitchfork bifurcation and a phase transition \footnote{the critical exponents of our model are however such that we cannot observe a super-linear scaling of the efficiency versus power \cite{Campisi2016_vol7}.} Furthermore, the dynamics and thermodynamics of the collective machine can be exactly mapped (at steady-state) from the many-body microscopic space into a much smaller density space \cite{SM}. Consequently, both the mean field and the finite but large $N$ properties of the machine are accessible via numerically exact calculations. 

Our central result is that the efficiency of our collective machine operating as a heat engine increases with the number of interacting machines. This occurs before and after the bifurcation and despite the fact that interaction at finite $N$ suppresses the tight coupling property of the individual machines. Remarkably, the macroscopic limit ($N$ very large) restores the tight coupling and enables the collective machine to reach the reversible efficiency. To our knowledge, this is the first time that an explicit mechanism is proposed to reach tight coupling in a macroscopic device made of interacting nanomachines. We also find that the interaction enables each particle to carry more energy, thus increasing the heat and work fluxes across the machine. Interestingly the most mechanical power is produced after the bifurcation, when a new stable branch appears, but before it becomes the dominant one because this new branch corresponds to a dud engine (i.e. a machine producing no work).

\begin{figure}
\centering
\includegraphics[width=8cm]{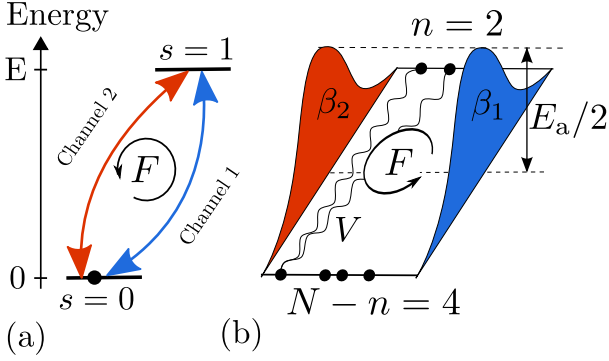}
\caption{
(a)~Single two state machine subjected to a non conservative force $F$ and which can change state due to two reservoirs. 
(b)~Ensemble of $N=6$ interacting machines in state $n = 2$. 
\label{FigModel}}
\end{figure}

\emph{Stochastic model and thermodynamics---} 
We start by considering a single noninteracting unicyclic nanomachine $i$, sketched in Fig.~\ref{FigModel}(a). It can be thought as a particle which can hop in two ways between a lower state $s_i=0$ of energy zero and an upper state $s_i=1$ of energy $E$. One way involves crossing an energy barrier of hight $E_\mathrm{a}$ by exchanging energy with the cold reservoir $\nu=1$ while another way involves crossing another energy barrier of the same hight but by exchanging energy with the hot reservoir $\nu=2$. Furthermore, hopping from $s_i=1$ to $s_i=0$ via channel $\nu=1$ requires to do work against the external nonconservative force $F$, while doing the same via channel $\nu=2$ gains work from $F$. The rate $k_{\epsilon}^{\nu} = \Gamma e^{-\frac{\beta_{\nu}}{2}\left[ E_\mathrm{a} + \epsilon E + \epsilon(-1)^\nu F \right]}$ therefore describes the probability per unit time for hopping upward ($\epsilon = +1$) or downward ($\epsilon=-1$) via channel $\nu$. $\Gamma = 1$ sets the time scale unit. 
In absence of force, the particle will in average move clockwise (i.e. go up via the hot reservoir and down via the cold one). When doing the same in presence of force, the machine operates as a heat engine which produces work by rotating against the force. When rotating on average counterclockwise (i.e. up via cold and down via hot reservoir), the machine operates as a heat pump since work is spent to bring energy from the cold to the hot reservoir.


We now turn to a collection of $N$ such unicyclic nanomachine as shown in Fig.\ref{FigModel}(b) interacting via an infinite range pairwise repulsive interaction of value $V/N$ between the particles with opposite states. The internal energy is thus
\begin{equation} \label{internal-energy}
U(\{s\}) \equiv E n + \f{V}{N} n(N-n) ,
\end{equation}
where $\{s\}$ denotes a many-body state of the collective machine and $n = \sum_{i=0}^N s_{i}$ the number of nanomachines in state $s_i=1$
\footnote{Via the mapping of state $s_i$ on the spin value $2 s_i -1 $, the internal energy of Eq.~\ref{internal-energy} is that of the infinite range Ising model with coupling constant $V/4$ and magnetic field $E/2$. We thus recover the Ising model when $\beta_{1} = \beta_{2}$ and $F=0$.}.  
Assuming that particle hop one at the time, the transition rate from $\{ s \}$ to $\{s\}_{i}^\epsilon$ due to reservoir $\nu$ reads
\begin{equation}\label{eq:rateMicro}
\omega^{(\nu)}_{\{s\}_i^\epsilon,\{s\}} \equiv 
\Gamma e^{-\frac{\beta_\nu}{2} \left[ E_\mathrm{a} + U(\{s\}_{i}^\epsilon)-U(\{s\}) +\epsilon(-1)^\nu F \right] },
\end{equation}
with $\{s\}_{i}^\epsilon = (s_{0},\dots,s_{i-1},(1+\epsilon)/2,s_{i+1},\dots,s_{N}) \neq \{ s \}$.

When the collective machine operates in a stationary state, its non-negative entropy production rate per nanomachine reads~\cite{VandenBroeck2014_vol418}
\begin{eqnarray} \label{eq:ExactEP}
\langle \sigma \rangle &\equiv& \frac{1}{N}\sum_{\{ s \}, \nu , i, \epsilon } \omega^{(\nu)}_{\{s\}_i^\epsilon,\{s\}} p^*(\{s \}) \ln \frac{\omega^{(\nu)}_{\{s\}_i^\epsilon,\{s\}} }{\omega^{(\nu)}_{\{s\},\{s\}_i^\epsilon} } \ge 0,
\end{eqnarray}
where $p^*(\{s \})$ is the stationary probability to find the system in state $\{s\}$. 
By substituting (\ref{eq:rateMicro}) in (\ref{eq:ExactEP}) as detailed in \cite{SM}, we find the more physically appealing decomposition $ \langle \sigma \rangle = \langle \sigma^\mathrm{w} \rangle + \langle \sigma^\mathrm{q} \rangle $,
where
\begin{equation}
\langle \sigma^\mathrm{w} \rangle = \beta_{1} \langle \dot{w} \rangle \equiv -2\beta_1 F \sum_{n=0}^{N-1} j_n^{(2)} , \label{eq:ExactWorkEP} 
\end{equation}
is proportional to $\langle \dot{w} \rangle$, the average work rate produced per machine, and $\langle \sigma^\mathrm{q} \rangle = (\beta_1-\beta_2) \langle \dot{q} \rangle $ is proportional to $\langle \dot{q} \rangle$, the heat rate per machine absorbed by the system from the hot reservoir. One has more precisely
\begin{eqnarray}
\langle \sigma^\mathrm{q} \rangle &=& \left( \beta_1-\beta_2\right) \sum_{n=0}^{N-1} \left[ V\left(1-\frac{2n}{N}\right) +E+F \right] j_n^{(2)}.
 \label{eq:ExactHeatEP}
\end{eqnarray}
In both Eqs.~(\ref{eq:ExactWorkEP}--\ref{eq:ExactHeatEP}), the net number of transitions per unit time from $n$ to $n+1$ due to reservoir $\nu$ is denoted
\begin{multline}
	N j^{(\nu)}_n \equiv \sum_{\{ s \} , i} \Big ( \omega^{(\nu)}_{\{s\}_i^1,\{s\}} \delta_{n,n(\{s\})} \\ - \omega^{(\nu)}_{\{s\}_i^{-1},\{s\}} \delta_{n+1,n(\{s\})} \Big) p^*(\{s \}).
\end{multline}
Kronecker's $\delta_{y,z}$ vanishes when $y \neq z$ and equals $1$ otherwise. From Eqs.~(\ref{eq:ExactWorkEP}--\ref{eq:ExactHeatEP}), we see that in absence of interactions, $V=0$, the property of tight coupling is satisfied. Indeed both the work and the heat rates are in this case proportional to the same current $\sum_{n=0}^{N-1} j_n^{(2)}$. However, this property is lost in presence of interaction since the heat looses this proportionality while the work does not.

Based on the entropy production decomposition (\ref{eq:ExactWorkEP}--\ref{eq:ExactHeatEP}), an unambiguous macroscopic efficiency of the machine operating as a heat engine ensues (see e.g. Ref.~\cite{Book_Bejan2006, Esposito2012_vol85a, Verley2014_vol5, Vroylandt2016_vol93})
\begin{equation}
 \eta \equiv - \frac{\langle \sigma^\mathrm{w} \rangle}{\langle \sigma^\mathrm{q} \rangle} = - \frac{\langle \dot{w}\rangle}{\langle \dot{q} \rangle} \frac{1}{\eta_\mathrm{rev}} \quad \text{with} \quad \eta_\mathrm{rev} = 1- \frac{T_1}{T_2}. \label{eq:ExactEff} 
\end{equation}
Indeed, in this case work is extracted, $\langle \dot{w} \rangle < 0$, heat is absorbed from the hot reservoir, $\langle \dot{q} \rangle > 0 $, particles rotates on average in the clockwise direction, and the efficiency is bounded by $1 \ge  \eta > 0$.
When $\langle \dot{w} \rangle > 0$ and $\langle \dot{q} \rangle < 0 $, the machine operates as a heat pump, particles rotate in the counter clockwise direction on average, and the macroscopic efficiency of the heat pump, $1/\eta$, is bounded by $1 \ge 1/  \eta > 0$. The dud engine regime occurs when $\eta < 0$.

\begin{figure}
\centering
\includegraphics[height=9cm]{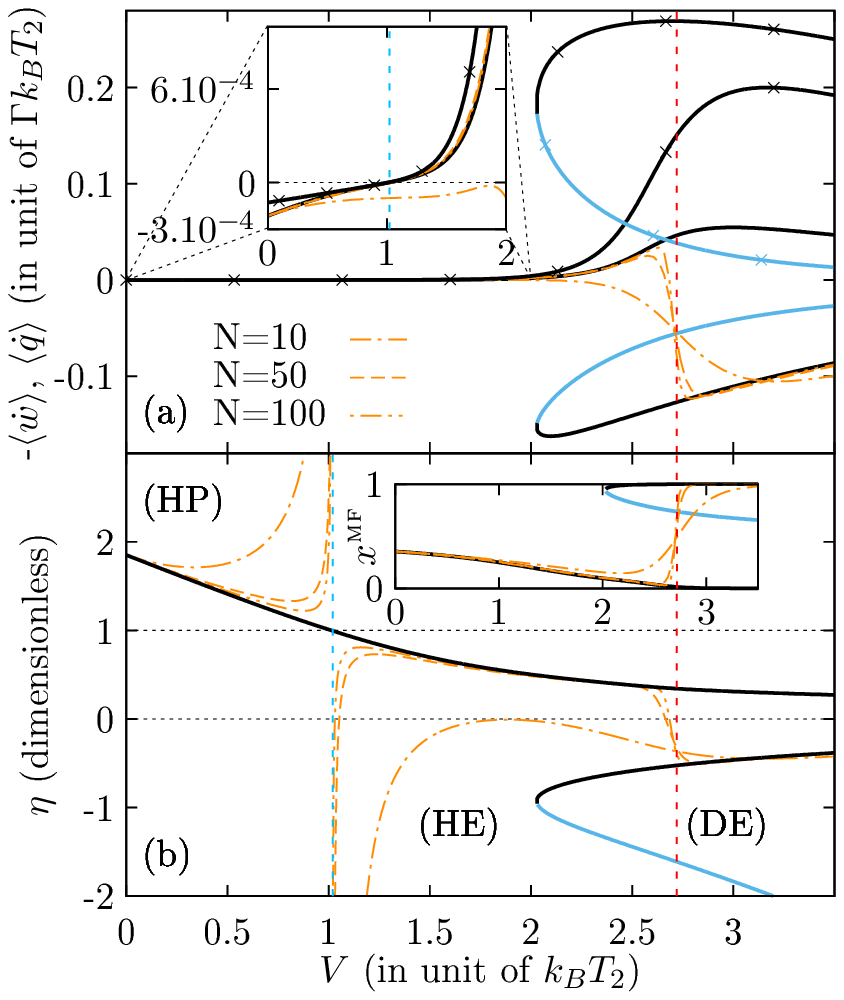} 
\caption{\label{FigMF} 
(a)~ Heat (resp. work) rate $\langle q \rangle$  (resp. -$\langle w \rangle$) per machine received (resp. delivered to the outside) by the ensemble of $N$ interacting machines, as a function of the interaction energy $V$. Crossed lines for $\langle q \rangle$ and solid lines for $\langle w \rangle$. Dashed lines correspond to output power for various values of $N$ and solid lines denote the stable (black) and unstable (light blue) mean field solutions ($N \to \infty$). HE, HP and DE denotes respectively the Heat Engine, the Heat Pump and the Dud Engine regimes.
Inset: Zoom of the input and output power for $V \in [0,2]$. 
(b)~ Macroscopic efficiency for finite $N$ (dashed lines) and in the mean field limit (solid lines). 
Inset: Stable (black) and unstable (light blue) mean field steady state densities $x^\MF$ as a function of the interaction energy $V$. 
The parameters are: $E_\mathrm{a}=2$, $E=0.1$, $\beta_1=10$, $\beta_2=1$, $F=0.5$.
}
\end{figure}

\emph{Mean field description---} We denote by $x\equiv n/N$ the density of particles in the upper state. One can first attempt to solve the master equation ruling the evolution of the probability $p(\{s \},t)$ of state $\{s\}$ at time $t$ by making use of a mean field approximation. The resulting nonlinear equation for the mean field density $ x^\MF $ reads :
\begin{equation}
\frac{d x^\MF }{d t} = \sum_{\epsilon,\nu}  \left ( \delta_{1,\epsilon}- x^\MF  \right ) e^{ -\frac{\beta_\nu}{2} \left[ E_\mathrm{a} + \epsilon V(1-2x^\MF)+\epsilon E +\epsilon(-1)^\nu F \right] }. \label{eq:MFdensity}
\end{equation}
The stationary solution of this equation is plotted in the inset of Fig.~\ref{FigMF}(b). We see that the density undergoes a bifurcation indicating a first order phase transition \cite{Cleuren2001_vol54}. 

We now turn to the mean field approximation for the heat and work parts of the entropy production that become
\begin{eqnarray}
 \sigma^{\mathrm{w}}_{\MF} & = & - 2\beta_1 F j_{N  x^{\MF}}^{(2)}, \label{eq:MFheatANDworkEPs1} \\
 \sigma^{\mathrm{q}}_{\MF} & = & \left(\beta_1-\beta_2\right) \left[  V(1-2 x^\MF) +E+F \right] j_{N  x^{\MF}}^{(2)}, 
\label{eq:MFheatANDworkEPs}
\end{eqnarray}
because the number of particle in the upper state converges to $N x^\MF$ in the macroscopic limit. Note that the mean field approximation restores the tight coupling property in presence of interaction as both the work and heat rates become proportional to $j^{(2)}_{Nx^{\MF}}$ in the macroscopic limit and hence proportional to each other. The efficiency becomes in the mean field description
\begin{equation}
 \eta^\MF = - \frac{\sigma^{\mathrm{w}}_{\MF}}{\sigma^{\mathrm{q}}_{\MF}} = - \frac{\dot{w}_{\MF}}{\dot{q}_{\MF}} \frac{1}{\eta_\mathrm{rev}}.
\label{eq:MFefficiency}
\end{equation}
Due to the tight coupling property one expects this efficiency to be higher than the efficiency of a finite ensemble of interacting machines. 

\emph{Results---} In order to verify the emergence of tight coupling predicted by the mean field theory in the macroscopic limit, we now numerically study the performance of the finite ensemble of $N$ interacting machines \cite{SM}. Fig.~\ref{FigMF}(a-b) depicts the work and heat rates and the efficiency as a function of $V$ for different values of $N$.
These results confirm that the finite $N$ calculations converge to the mean field result as $N$ is increased. They also verify that the efficiency is higher in the macroscopic limit than at finite $N$. Without interaction ($V=0$), the machines behave as a heat pump; as the interaction is increased, the heat pump becomes more efficient, since $1/ \eta$ increases. Given that the mean field machine displays tight coupling, the operating mode switches from the heat pump to the heat engine regime at the reversible efficiency $\eta^{\MF} = 1$ which corresponds to equilibrium. Using Eqs.~(\ref{eq:MFdensity}) to (\ref{eq:MFefficiency}), we can predict that the switch occurs at $V=1.02$ for the set of parameters used in Fig.~\ref{FigMF}(a-b). A striking feature is that right above (resp. below) this value, the efficiency of the finite $N$ heat engine (resp. heat pump) drops dramatically. For values of $V$ right above $V=1.02$, the interacting machine is even briefly dud before quickly coming back to a heat engine regime. This singular behavior is due to the lack of tight coupling between the heat and work rates. Indeed, when the heat received from the hot reservoir vanishes, $\eta$ diverges since the work can take a finite value in absence of tight coupling, as shown in the inset of Fig.\ref{FigMF}(a). Instead, when tight coupling is restored in the large $N$ limit (i.e. at the mean field level), both work and heat vanish together (even in presence of a finite temperature gradient and force) while the efficiency involving their ratio tends to one. This would be impossible without tight coupling, making non-tightly coupled machines systematically more dissipative. As the interaction is further increased, the efficiency of the heat engine starts to decrease while the work rate is significantly increased. When the interaction reaches the critical value located at the (vertical) dashed red line $V=2.72$, a first order phase transition occurs which makes the machine dud. The work rate and efficiency of the finite $N$ interacting machines (orange dashed lines) suddenly drops because the systems moves from the old stable branch corresponding to a heat engine regime to another one corresponding to a dud regime. Both branches are denoted by black solid lines and the transition from one to another is clearly seen on the finite $N$ unique solution.

\begin{figure}
\centering
\includegraphics[height=9cm]{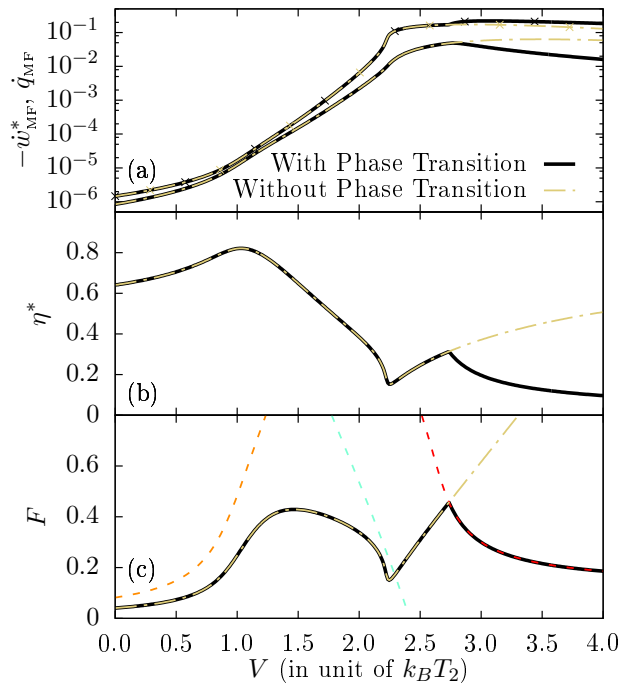}
\caption{\label{figMP} 
(a)~Mean field work rate per machines maximized with respect to the force $F$ as a function of the interaction energy $V$ (lines without crosses) and corresponding mean field heat rate (lines with crosses).
The black lines correspond to the work rate optimization when the system undergoes the phase transition. The beige dot-dashed lines are obtained when the optimization is performed by forcing the system to remain on the same branch before and after the transition (i.e. one artificially suppresses the phase transition). 
(b)~Efficiency at maximum work rate as a function of the interaction $V$.  
(c) Force maximizing the work rate, $F^*$. 
The three orange, light blue and red dashed lines (from left to right) denote respectively transition lines in $F$ above which the machine operates as a HP and below which it operates as a HE (orange), above which multistability emerges (light blue), above which the phase transition occurs (red).
Other parameters: $E_\mathrm{a}=2.0$, $E=0.1$, $\beta_1=10$, $\beta_2=1$. The heat and work rates are in units of $\Gamma k_\mathrm{B} T_{2}$ and $F$ is in unit of $k_\mathrm{B} T_{2}$. }
\end{figure}

In Fig.~\ref{figMP}(a), we consider the mean field work and heat rate per machine as a function of the interaction $V$ when the work rate is maximized with respect to the force $F$. 
We clearly see that as the interaction is increased, up to a five order of magnitude growth in the work rate delivered per machine is observed. This enhancement persists as long as the phase transition has not occurred. Beyond this point, the work rate starts decreasing. The heat rate follows a similar trend but saturates instead of decreasing after the phase transition.
The corresponding efficiency at maximum power, $\eta^*$, is represented on Fig.~\ref{figMP}(b). It follows a trend similar to the value of the force which maximizes the work rate, $F^{*}$, and which is represented in Fig.~\ref{figMP}(c). Both curves display two maxima separated by a same minimum. The second maximum is very abrupt and corresponds to the phase transition. Interestingly, after this second maximum, $F^{*}$ starts following the red-dashed critical line (i.e the critical value of $F$ at which the transition occurs for a given $V$). The line is not crossed by the optimization procedure because for greater values of the force, the phase transition would push the machine into the new stable branch which produces less power. The loss in power and efficiency after the second maximum can thus be seen as the price to pay for preventing the phase transition to occur.  

\emph{Conclusions---} By studying power generation and its efficiency using an explicit model of interacting machines undergoing a phase transition, we were able to draw two main conclusions: interactions between a large number of machines can respectively enhance the power generation and the efficiency. 
Further insight might be revealed by studying efficiency fluctuations \cite{Verley2014_vol5, Verley2014_vol90, Gingrich2014_vol16, Martinez2015_vol, Proesmans2015_vol17, Proesmans2015_vol92, Vroylandt2016_vol93, Proesmans2016_vol6}.
The emergence of tight coupling in the thermodynamic limit can be seen as resulting from the emergence of a conservation law. Indeed, it was recently show in Ref. \cite{Polettini2016_vol94} that the number of independent thermodynamic forces controlling the steady state entropy production of a machine is equal to the number of thermodynamic intensive variable characterizing the reservoirs, here three ($\beta_1$, $\beta_2$ and $F$), minus the number of conservation laws (i.e. the number of constrains between steady state currents). In absence of tight coupling this number is one due to energy conservation in the system and as a result two independent forces ensue: $\beta_1-\beta_2$ and $\beta_1 F$. But tight coupling, by further constraining the currents, creates an additional conservation law which results in a single independent force instead of two. This latter is easily obtained as the prefactor of the current when summing (\ref{eq:MFheatANDworkEPs1}) and (\ref{eq:MFheatANDworkEPs}).
The present model provides an explicit mechanism demonstrating that new conservation laws can emerge in the thermodynamic limit.
The generality of this mechanism is still to be better understood and further investigations are required to determine if a similar mechanism can exist for machines modeled by more complex graphs or for ensemble of machines with short range interactions \cite{Cornu2017_vol2017}. In any case, our results provide an interesting hint on how to design highly efficient machines producing significant power.


\textit{Acknowledgment.---}
We thank C. Van den Broeck for interesting discussions during the early stage of this work. 
This research was funded by the National Research Fund Luxembourg (project FNR/A11/02 and INTER/FWO/13/09) and by the European Research Council (project 681456).

\bibliography{Ma_base_de_papier}
\end{document}


\title{Supplementary Material: Collective effects enhancing power and efficiency}
\author{Hadrien Vroylandt}
\affiliation{Laboratoire de Physique Th\'eorique (UMR8627), CNRS, Univ. Paris-Sud, Universit\'e Paris-Saclay, 91405 Orsay, France}
\author{Massimiliano Esposito}
\affiliation{Complex Systems and Statistical Mechanics, Physics and Material Science Research Unit, University of Luxembourg, L-1511 Luxembourg, G.D. Luxembourg}
\author{Gatien Verley}
\affiliation{Laboratoire de Physique Th\'eorique (UMR8627), CNRS, Univ. Paris-Sud, Universit\'e Paris-Saclay, 91405 Orsay, France}

\date{\today}

\maketitle

\renewcommand\theequation{M\arabic{equation}}
\renewcommand\thefigure{F\arabic{figure}}
\setcounter{equation}{0}
\setcounter{figure}{0}

In this supplementary material, we show that the dynamics of the many-body system can be coarse-grained exactly. Then, we prove that the entropy production can also be coarse-grained exactly. Finally, we derive the explicit stationary probability enabling to compute the heat and work rates of the collective machine with a finite number of particles.\\

The exact dynamics in term of microscopic states (i.e. many-body states), $\{s\}$, introduced in the letter in Eq.~(\ref{eq:rateMicro}), can be exactly mapped into a dynamics on mesostates $n(\{ s\}) \equiv \sum_{i=0}^{N} s_i$ denoting the number of particles in the upper state. The mesostate probability $p(n,t) =  \sum_{\{s\}} p(\{s \},t) \delta_{n(\{s\}),n} $ evolves according to
\begin{equation}\label{eq:MasterEqMeso}
\Dp{}{t}p(n,t) = \sum_{\epsilon=\pm 1}  p(n+\epsilon ,t) k_{n+\epsilon,n}  - p(n,t) \sum_{\epsilon=\pm 1} k_{n,n+\epsilon},
\end{equation}
where the transition rates for jumping from $n \rightarrow n+\epsilon$ due to reservoir $\nu$ are given by
\begin{eqnarray}
k^{(\nu)}_{n+\epsilon,n} &=& \sum_{i} \omega^{(\nu)}_{\{s\}^\epsilon_i,\{s\}} \delta_{n,n(\{s\})} \delta_{s_i+\epsilon,(\epsilon+1)/2} , \\
&=& N \left ( \frac{1+\epsilon}{2 } -\epsilon  \frac{n}{N} \right ) e^{ -\frac{\beta_\nu}{2} \left( E_\mathrm{a} + \epsilon V(1-2\frac{n}{N})+\epsilon E +\epsilon(-1)^\nu F \right) }. \nonumber
\end{eqnarray}
This result is due to the fact that the microscopic rates in Eq.~(\ref{eq:rateMicro}) are the same for all microstates $\{s\}$ associated to the same mesostate $n$. The mesoscopic rates satisfy the local detailed balance 
\begin{equation}
\ln \frac{k^{(\nu)}_{n+\epsilon,n}}{k^{(\nu)}_{n, n+\epsilon}} = -\beta_\nu (\mathcal{F}_{n+\epsilon}^{(\nu)}-\mathcal{F}_{n}^{(\nu)} - W^{(\nu)}_{n+\epsilon,n}) \label{eq:LDB},
\end{equation}
where each state has now an associated free energy $\mathcal{F}_n^{(\nu)}=U_n-S_n/\beta_{\nu}$ with an energy $U_n= Vn(N-n)/N$ and an internal entropy $S_n=\ln N!/[n!(N-n)!]$. The elementary work $W^{(\nu)}_{n+\epsilon,n}= \epsilon(-1)^\nu F$ represents the energy provided by the non conservative force at each jump. The total rates $k_{n+\epsilon,n} = \sum_\nu k^{(\nu)}_{n+\epsilon,n}$ are not detailed balance. \\

\begin{figure}
\centering
\includegraphics[width=8cm]{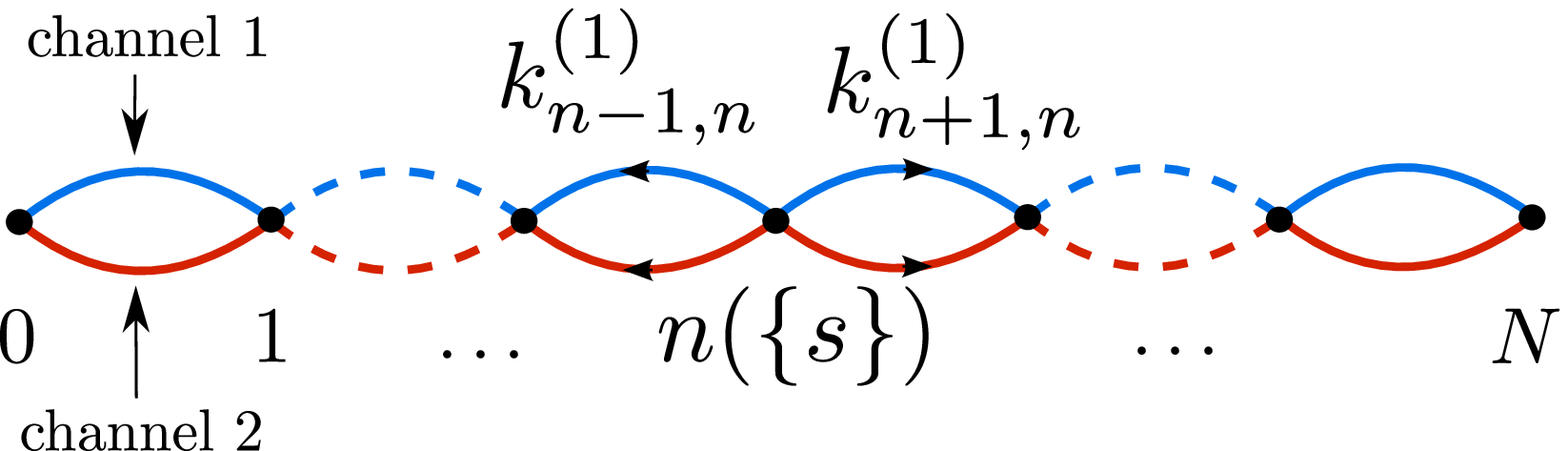}
\caption{
Network representation of an ensemble of $N$ interacting machines; the two types of edges correspond to the hot (red) and cold (blue) heat reservoir. Indiscernibility allows us to identify all the states $\{s\}$ with the same number of particles $n=n(\{s\})$ in the upper state.
\label{FigNetwork}}
\end{figure}

Using Eqs.~(\ref{eq:rateMicro}) and (\ref{eq:ExactEP}) of the main text, the entropy production rate can be rewritten as
\begin{eqnarray} \nonumber
 \langle \sigma \rangle &=& -\frac{1}{N}\sum_{\{ s \}, i, \epsilon } \omega^{(1)}_{\{s\}_i^\epsilon,\{s\}} p^*(\{s \}) \beta_1 \left[ U(\{s\}_{i}^\epsilon)-U(\{s\}) - \epsilon F \right], \nonumber \\ 
 &&- \frac{1}{N}\sum_{\{ s \}, i, \epsilon } \omega^{(2)}_{\{s\}_i^\epsilon,\{s\}} p^*(\{s \}) \beta_2 \left[ U(\{s\}_{i}^\epsilon)-U(\{s\}) + \epsilon F \right], \nonumber \\
 \langle \sigma \rangle &=& \frac{2\beta_1 F}{N}\sum_{\{ s \}, i, \epsilon } \epsilon \omega^{(1)}_{\{s\}_i^\epsilon,\{s\}} p^*(\{s \}) - \frac{1}{N}\sum_{\{ s \}, i, \epsilon } p^*(\{s \})  \left[ U(\{s\}_{i}^\epsilon)-U(\{s\}) + \epsilon F \right] \sum_\nu\beta_\nu \omega^{(\nu)}_{\{s\}_i^\epsilon,\{s\}} , \nonumber \\
 \langle \sigma \rangle & = & 2\beta_1 F \sum_{n} j_n^{(1)}- \frac{1}{N} \sum_{n,\nu} \beta_\nu  \sum_{\{s\},i,\epsilon} \epsilon \left[ V\left(1-\frac{2n}{N}\right) +E+F \right]  \omega^{(\nu)}_{\{s\}_i^\epsilon,\{s\}} p^*(\{s \})  \delta_{n+(1-\epsilon)/2,n(\{s\})} , \nonumber \\
 \langle \sigma \rangle & = & 2\beta_1 F \sum_{n} j_n^{(1)}-  \sum_{n,\nu}   \left[ V\left(1-\frac{2n}{N}\right) +E+F \right] \beta_\nu  j^{(\nu)}_{n} =  \langle \sigma^{\mathrm{w}} \rangle +  \langle \sigma^{\mathrm{q}} \rangle, \label{eq:SM-ExactEP} 
\end{eqnarray}
which is the result obtained in Eqs.~(\ref{eq:ExactWorkEP}) and (\ref{eq:ExactHeatEP}) when using $j^{1}_{n}+j^{2}_{n} = 0$. Indeed, the total probability current to the right should vanish in the stationary state, implying $j^1_n + j^2_n = 0$ for all $n$, as is clear from Fig.~\ref{FigNetwork}. 

We now turn to the stationary probability given by the spanning tree formula~\cite{Schnakenberg1976_vol48}:
\begin{equation}
p\e{stat}(n) \propto \sum_{T_\alpha(n)} \prod_{ (n,\epsilon) \in T_\alpha(n)} \sum_{\nu} k^{(\nu)}_{n+\epsilon,n}.
\end{equation}
The sum runs on all spanning trees $T_\alpha(n)$ rooted in $n$.
The product spans all possible edges (oriented to the root) in a tree: $(n,\epsilon)$ is the edge associated to the transition $n \rightarrow n+\epsilon$. For the network displayed in Fig.~\ref{FigNetwork}, the sum on spanning trees can be factorized into the more explicit expression
\begin{eqnarray}
p\e{stat}(n) &=& \displaystyle \frac{1}{ Z} \left[ \prod_{m =0}^{n-1}\sum_{\nu} k^{(\nu)}_{m+1,m} \right] \left[ \prod_{m=n+1}^N \sum_{\nu} k^{(\nu)}_{m-1,m} \right ], \label{eq:ProbaStat} 
\end{eqnarray}
where $ Z$ is a normalization constant scaling like $N^{N}$. 
Using this stationary probability, the steady state probability currents read 
\begin{equation} \label{eq:stationary-prob-current}
N j^{(2)}_{n} = k^{(2)}_{n+1,n} p\e{stat}(n)  -  k^{(2)}_{n,n+1} p\e{stat}(n+1).
\end{equation}
The finite size results of Fig.~\ref{FigMF} are obtained using Eqs.~(\ref{eq:SM-ExactEP}), (\ref{eq:ProbaStat}) and (\ref{eq:stationary-prob-current}).

\bibliography{Lien_ma_base_de_papier}